\documentstyle[referee,psfig]{aa}

\def\R23{\mbox{$\rm R_{23}$}}
\def\lya{\mbox{Ly-$\alpha$}}
\def\am{\mbox{\AA\,}}
\def\arcmin{\hbox{$^\prime$}}
\def\arcsec{\hbox{$^{\prime\prime}$}}
\def\degr{\hbox{$^\circ$}}

\def\kmsmpc{km s$^{-1}$ Mpc$^{-1}$}

\def\Hb{\mbox{${\rm H}{\beta}$}}
\def\Ha{\mbox{${\rm H}{\alpha}$}}
\def\OIIIa{\mbox{${\rm [O\,III]\,}{\lambda\,5007}$}}

\def\OII{\mbox{${\rm [O\,II]\,}{\lambda\,3727}$}}

\begin{document}
\title{Constraints to the evolution of  Ly-$\alpha$ bright galaxies
  between $\rm{z}=3$ and $\rm{z}=6$\thanks{Based on observations
  obtained at the ESO VLT, Paranal, Chile; ESO programs 67.A-0175 and
  68.B-0088}}

\author{C. Maier\inst{1}, K. Meisenheimer\inst{1},
 E. Thommes\inst{1,2},
 H. Hippelein\inst{1},
  H.~J. R\"oser\inst{1},\\ J. Fried\inst{1}, B. von Kuhlmann\inst{1},
  S. Phleps\inst{1}, C. Wolf\inst{1,3}}

\offprints{C. Maier, \email{maier@mpia-hd.mpg.de}}

\institute{Max--Planck--Institut f\"ur Astronomie, K\"onigstuhl 17, 
           D-69117 Heidelberg, Germany
\and Insitut f\"ur Theoretische Physik, Universit\"at Heidelberg,
           69111 Heideleberg, Germany 
\and Department of Physics, Denys Wilkinson Bldg., University of
           Oxford, Keble Road, Oxford, OX1 3RH, U.K.
}

\date{Received 6 November 2002 / Accepted 6 February 2003}

\abstract{
Galaxies at high redshift with a strong  Ly-$\alpha$ emission line trace
massive star formation in the absence of dust, and can therefore be
regarded as a prime signature of the first major starburst in galaxies.
We report  results of the \lya\, search within the Calar Alto Deep Imaging Survey (CADIS).
With imaging  Fabry-Perot interferometer
CADIS can
detect emission lines in three waveband windows free of night-sky
emission  lines
at 700\,nm, 820\,nm, and 920\,nm. The typical flux detection limit for \lya\,
emission redshifted into these windows,  $F_{lim} \ga 3
\times 10^{-20} \rm{W}\,\rm{m}^{-2}$, corresponds to  (unobscured) star
formation rates of  $\ga 10\,\rm{M}_{\odot}$/yr at $z=6$.
Candidate Ly-$\alpha$-emitting galaxies
are selected from the total emission line sample, which contains more than
97\% of objects at $z<1.2$,
by the absence of flux below the Lyman limit (B-band
``dropouts''), and the non-detection of secondary emission  lines in  narrow band  filters.
We have detected  5 bright  \lya-emitting galaxy candidates at
$z \simeq 4.8$, and 11 candidates  at $z \simeq 5.7$.
For two of four observed Ly-$\alpha$ candidates, one candidate
  at $z
  \simeq 4.8$, and the other at $z \simeq 5.7$,
the emission line  detected with the
Fabry-Perot has been verified spectroscopically at the VLT.
%
When compared to \lya\, surveys at $z\leq3.5$ even the upper limits
set by our list of candidates show that bright \lya\, galaxies are
significantly rarer at $z\ga 5$ than the assumption of a non-evolving population would predict.
Therefore we conclude that the \lya\, bright phase of primeval
star formation episodes  reached its peak at redshifts $3<z<6$.
}

\titlerunning{Abundances of bright   Ly-$\alpha$  galaxies at $z>3$}
\authorrunning{Maier et al.}
\maketitle

%

\section{Introduction}

Important observational information about the  epoch of galaxy
formation can only be gained from  the
existence, number counts and properties  of galaxies at very high redshift.
In the past couple of years substantial progress has been made in
discovering galaxies at redshifts $z>4$.
Several dozens of Lyman break galaxies have been detected at $z \sim
4$ (e.g., Steidel et al. \cite{steidel99}).
In addition, a substantial number of
galaxies at $z>5$ have been detected and verified
spectroscopically (e.g., Dey et al. \cite{dey}; Weyman  et al.  \cite{weyman}; van Breughel et al. \cite{vanBr};
Hu et al.  \cite{hu99}; Hu et al.  \cite{hu02}; Ellis et al.
\cite{ellis}; Dawson et al. \cite{dawson}; Ajiki et
al. \cite{ajiki}; and Rhoads et al. \cite{rhoads02}); one of
these (van Breughel et al. \cite{vanBr}) is a radio galaxy at $z=5.2$. 
Quasars have been identified out to $z = 6.3$ (e.g., Pentericci 
et al. \cite{pent}; Fan et al. \cite{fan01}).
However,
there is a lack of \emph{systematic} surveys for high redshift
primeval galaxies. Many of the
galaxies at  $z>5$ have been detected serendipitously.

Moreover, as it has been pointed out, e.g., by Steidel et
al. (\cite{steidel96}),  the strong metal absorption lines in the
spectra  of Lyman break galaxies indicate that they
have already formed
at least one generation of stars which has chemically
enriched the systems and lead to dust formation, making Ly-$\alpha $
emission weak or undetectable.
Therefore, the Lyman break galaxies do not show  the properties expected for  primeval galaxies.
In contrast to the Lyman break galaxies, all but one of the  galaxies found at $z > 5$
exhibit a very strong \lya\, emission line, in agreement with  
population synthesis models for primeval galaxies of Charlot \& Fall (\cite{charlot}), which predict
that a young, dust free, star forming galaxy
should show strong Ly-$\alpha $ emission with (intrinsic) equivalent widths
in the range of 5-25\,nm.

A direct probe of galaxy formation is to determine the number counts
and redshift evolution of those nascent galaxies.
Several systematic surveys aimed at discovering a significant number
of $z >  4.5$ galaxies have now been proposed and are being carried
out, e.g.,  Cowie \& Hu (\cite{cowie}), Rhoads et al. (\cite{rhoads})
and our survey, the \textbf{C}alar \textbf{A}lto \textbf{D}eep \textbf{I}maging
 \textbf{S}urvey (CADIS, Meisenheimer et al. \cite{meise98}, \cite{meise03}).
At $z=4.5$, a universe with standard parameters $H_{0}=70$ \kmsmpc, $\Omega_{M}=0.3$, $\Omega_{\Lambda}=0.7$   is only 1.2 Gyr
old, corresponding to a look-back time of $>90$\% of the age of the
universe.
Any object observed at this or an earlier epoch should mark the early
stages of galaxy formation.

Unlike the Lyman break technique (e.g., Steidel \& Hamilton
\cite{steidel92}, \cite{steidel93}), which
selects galaxies by the presence of young stars and intervening HI absorption,
CADIS selects galaxies by the  extreme equivalent widths of their
Lyman-$\alpha$-emission ($W_{\lambda_{rest}} > 10\,\rm{nm}$).
It has been demonstrated by Kudritzki et al. (\cite{kudr}) that the high Ly-$\alpha$ equivalent width of these galaxies
can only arise in  regions nearly free of dust, and therefore the \lya\,
emission line is a good
tracer of star formation from primordial material.
Unlike similar
programmes at 8$-$10 m telescopes, CADIS aims
to find \emph{the most luminous} starbursts, with SFR $\geq 10\,
\rm{M}_{\odot}$/year, which most likely led  to the formation
of the galactic bulge. Even in the age of 10-m telescopes, only these
luminous starbursts are accessible for detailed spectroscopic
follow-up.

This paper is structured  in the following way: In section 2 we
describe how we select \lya\, galaxy candidates at $z>4.7$ with CADIS.
In section 3 we present the spectroscopic follow-up of some \lya\, candidates.
Finally, in section 4  we discuss the comparison between observed and theoretical
number counts of \lya-emitting galaxies up to $z \approx 6$.


\section{Lyman-$\alpha$ candidates from the CADIS survey}

CADIS
 combines  a moderately deep
multi-band survey (10 $\sigma$ limit, R$_{\rm{lim}} = 24$) with a deep
emission line survey employing an imaging
Fabry-Perot-Interferometer ($\rm{F}_{\rm{lim}} = 3 \cdot 10 ^ {-20}
 \rm{W}  \rm{m}^{-2}$), which covers
three waveband windows essentially free from  atmospheric OH emission lines:  A\,($\lambda = 702 \pm 6$\,nm),
B\,($\lambda = 819 \pm 5$\,nm), and C\,($\lambda = 920 \pm 8$\,nm).
The classification and analysis of emission line galaxies from CADIS
 are outlined in Hippelein et al. (\cite{hippe}), Meisenheimer et
 al. (\cite{meise03}).
Therefore, in this section we will present  only those aspects which are relevant to
 the search of \lya\, galaxies using the CADIS (emission line) survey.

\begin{figure}[ht!]               
\centerline{
\psfig{figure=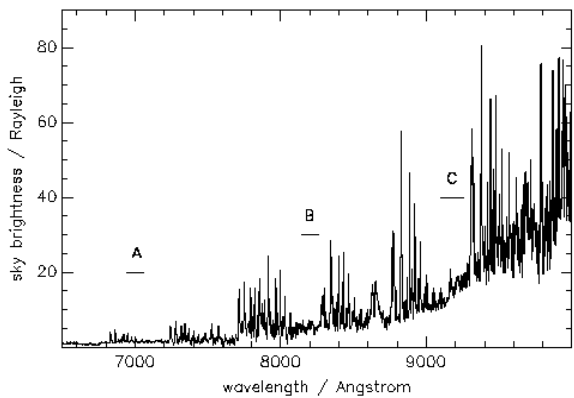,clip=t,width=8.0cm,angle=0}}
\caption[]{\label{nightemis}\footnotesize Night-sky emission spectra. The windows A, B, and C which are
  searched by CADIS are essentially free from atmospheric OH emission lines.}
\end{figure}

The optimal detection of a weak emission line galaxy superimposed on
the bright night sky is reached when the instrumental resolution
$\delta \lambda$ is adapted to the expected line width. Assuming
$\Delta v \approx 300\,\rm{km}\,\rm{s}^{-1}$, a typical  virial
velocity of a Milky Way-like galaxy, yields  $\delta
\lambda \cong  (1+z) (\Delta v/c) \lambda_{0} \cong 0.9$\,nm for
a Lyman-$\alpha$ emission line at $z \sim 6$.
On the other hand, the detection probability increases with the
observed volume $\Delta V \sim  \Delta z \Delta \Omega \sim  \Delta
\lambda \Delta \Omega$, where $\Delta \Omega$ denotes the solid angle surveyed.
Using a Fabry-Perot (FP) Etalon with $\Delta \lambda \cong 2$\,nm,
which is tuned in 9 steps and covers an interval of
$12-15$\,nm, CADIS  provides an optimal compromise of both   sensitivity and
search volume.
The three wavelength intervals (A, B, and C, see Figure \ref{nightemis}) which are scanned by CADIS
correspond to Lyman-$\alpha$ redshifts
 of $z \cong 4.8$, 5.7, and 6.6, respectively. The field size is $\Delta \Omega
 \approx 100$ arcmin$^{2}$.
 Model predictions 
(Thommes \& Meisenheimer 1995)
  of the galaxy abundance down to a line flux limit
 F$_{\rm{line}} \geq 3.0 \cdot 10^{-20}\,\rm{Wm}^{-2}$ (which can be reached with the
 Calar Alto 2.2- and 3.5-m telescopes within a few hours of observing time) made clear,
 before the project had  even started, that at most a handful of
 Lyman-$\alpha$ galaxies would be expected per field and wavelength interval
 in the case of an open universe,  which is commonly assumed today
 ($\Omega_{M}=0.3$, $\Omega_{\Lambda}=0.7$). 
 Accordingly  we decided to search every wavelength interval in five fields.
After six years of observations on Calar Alto, 90\% of the planned
Fabry-Perot data have been obtained.
The Fabry-Perot wavelength intervals which have been almost completely
analysed are given in Table \ref{tabfields}.
The analysis of interval C has been postponed for the moment because of technical
problems (strong ``fringing'' in the  CCD images), and low
expectations of success on the basis of the present results at $z<6$,
which will be presented in this paper.

\begin{table}[ht!]
\caption[CADIS fields]{\label{fields} \footnotesize The analysed
  FP-windows   in four CADIS fields.}
\vspace{1ex}
\label{tabfields}
\begin{center}
\begin{tabular}
{    c       c                c                  c              c  c}
\hline
   Field  & FP-window   & \multicolumn{2}{c}{Center coordinates} &
   $\Delta \Omega $ & $ \Delta z$ \\
\hline   
  name    & observed &      $\alpha$(2000)  & $\delta$(2000)  &
   arcmin$^{2}$ & analysed
   \\
          &          &        ($h\,m\,s$)     &  (\degr\,  \arcmin\, \arcsec)              &  &        \\
 \hline   &          &                      &                 &          \\[-4mm] 
   01h    &   B    & 01 47 33.3 &   02 19 55    &    105   & 0.095    \\
   09h    &   A   &  09 13 47.5 &   46 14 20    &    98   & 0.082 \\
          &   B   &  &   &      & 0.070 \\
   16h    &   B   &   16 24 32.3 &   55 44 32    &    107 & 0.095 \\
   23h    &   A   &  23 15 46.9 &   11 27 00    &    103  &  0.057  \\
          &   B   &  &  &      &  0.082  \\
\hline  
\end{tabular}
\end{center}
\end{table}

The following results are based on the data in the
FP-windows   in four
CADIS fields presented in Table \ref{tabfields}.
It should be noted that the analysis of 69\% of  window B, and \emph{only} 27\% of window A has been
finished  (see Table \ref{fields}), in terms of the final $\Delta
\Omega \times \Delta z$: $\Delta \Omega_{final}$ will be $\approx 500\,
\rm{arcmin}^{2}$, corresponding to five CADIS fields,
 $\Delta z_{final}$ for each Fabry-Perot window  will be $\Delta
z_{final} \approx 0.1$. 

The current classification of emission line galaxies shows that more
than  97\% 
of the  emission line galaxies are foreground galaxies ($z
< 1.2$), in which we mainly detect   the lines \Ha, \Hb, \OIIIa, and
\OII.
Therefore, the main challenge is to separate those foreground objects
from the \lya\, candidates.
We identify  Lyman-$\alpha$-galaxy candidates among the emission
line galaxies  by the following
exclusion criteria:

(1)
There should be no flux below the Lyman limit of a
Lyman-$\alpha$ galaxy, because of an intrinsic drop in the spectra of hot O and B stars at
the Ly-limit (Charlot \& Fall \cite{charlot}), and absorption by the
neutral interstellar medium either in the galaxy
itself (Leitherer et al. \cite{leith95a}, \cite{leith95b}) or in
Lyman-limit foreground systems close to the primeval galaxy.
For $z > 4.7$ the Lyman limit lies at $\lambda > 520$\,nm.  Therefore,
no flux should be detected in the CADIS B filter
($\lambda_{cent}=461$\,nm,  FWHM$=113$\,nm). Accordingly,  we require that
$F_{B}<2 \cdot \hat{\sigma}_{B}$ (``blue dropouts''), where $F_{B}$ is the flux measured in the
B filter, and $\hat{\sigma}_{B}$ its true overall error which takes
into account both  calibration errors and the scatter of the flux
between the individual images.
From 614 emission line galaxies selected in four fields in the
FP-windows given in Table \ref{fields}  this
 criterion yields
 $\rm{N}_{\rm{Noblue}}= 85$ galaxies, i.e. 13.8\% of the 614 emission line galaxies.

(2) 
Every single remaining \lya\, candidate has to be carefully checked on
the FP, pre-filter and R images because of the possible contamination
by nearby bright objects. Candidates which are closer than about
3\arcsec\, to a bright object are most likely spurious,  since the  changing atmospheric conditions change
the tails of bright objects affecting the photometry of close-by
galaxies. Therefore, such spurious objects are removed from the
list of candidates. 
After this step, 5.7\% of the emission line galaxies remain.

(3)
Foreground galaxies, for which we detect one of the  prominent
emission lines \Ha, \Hb, \OIIIa, or \OII\, in the Fabry-Perot, can be
excluded as soon a secondary line is detected in one of the    veto filters,
 leaving  4.7\% of the galaxies.

(4)
The distinction between a (rare) galaxy with bright \OII\,
emission but weak restframe UV continuum, i.e., undetectable in the
CADIS B filter, and a Lyman-$\alpha$-galaxy is
ambiguous based on
the CADIS veto filters, because no line bluewards of \OII\, can be
detected. Instead, we have to use the entire spectrum for this decision.
Lyman-$\alpha$ galaxies at high redshift should distinguish themselves by a continuum step
across the Lyman-$\alpha$ line, due to  absorption from neutral hydrogen in the
Ly-$\alpha $ clouds and Ly-limit systems along the line of sight
(Madau  \cite{madau}).
Therefore, the candidates which remain after step (3) are classified
in two
categories: likely \lya, if they show almost no continuum on the
blue side of the \lya\, emission line, e.g., no significant flux in
the R filter, and likely \OII\, objects otherwise.
However, we need  spectroscopic follow up observations of candidates
belonging to \emph{both} categories, since it cannot be excluded for sure,
that some \lya\, galaxies fall between the  likely \OII\, objects.
 After
this step,
16 Lyman-$\alpha$-candidates (classified as most likely \lya) remain
(see Table \ref{listlyman}),
i.e., 2.6\% of the emission line galaxies we
found in four CADIS fields.
In addition we are left with 13 likely \OII\, candidates.

\begin{table}[ht!]
  \caption
[The current list of  16 CADIS Lyman-$\alpha$-candidates]
{\label{listlyman} \footnotesize The current list of  16 CADIS
  Lyman-$\alpha$-candidates.

 $^{v}$ The line of the \lya\, candidate  has  been verified by
 spectroscopic follow-up.
 
 $^{n}$ The line of the \lya\, candidate  has \emph{not} been verified by
 spectroscopic follow-up.

 $^{q}$ Probable  quasar. 
}
\begin{center}
\begin{tabular}{c c l c c }
\hline
FPI & Field &   Nr &z & $\rm{F}_{\rm{line}}(\rm{W}\rm{m}^{-2})$\\
\hline
A
&09h& 21556& 4.734 & 5.7$\cdot10^{-20}$  \\
&23h & 34751& 4.772 & 5.0$\cdot10^{-20}$ \\
&23h & 34105& 4.793 & 4.2$\cdot10^{-20}$ \\
&23h & 50707$^{v}$& 4.801 & 4.0$\cdot10^{-20}$ \\
&09h & 38114& 4.741 & 3.7$\cdot10^{-20}$ \\
\hline
B
&23h & 40663 & 5.732 & 6.4$\cdot10^{-20}$ \\
&23h & 23836 & 5.705 & 5.7$\cdot10^{-20}$ \\
&23h & 28548 & 5.694 & 5.6$\cdot10^{-20}$ \\
&23h & 45745 & 5.733 & 5.1$\cdot10^{-20}$ \\
&23h & 9324 & 5.730 & 4.8$\cdot10^{-20}$ \\
&16h &  3171 & 5.746 & 4.3$\cdot10^{-20}$ \\
&01h &  3238$^{v}$& 5.732 & 4.1$\cdot10^{-20}$ \\
&23h & 45065 & 5.735 & 4.1$\cdot10^{-20}$ \\
&16h &  2314$^{q}$& 5.694 & 3.4$\cdot10^{-20}$ \\
&01h &  27927$^{n}$& 5.677 & 3.1$\cdot10^{-20}$ \\
&01h &  28090$^{n}$& 5.681 & 2.9$\cdot10^{-20}$\\
\hline
\end{tabular}
\end{center}
\end{table}


\section{Spectroscopic observations}

Due to the low abundance of \lya\, galaxies (see section 4), spectroscopic follow-up at large telescopes
constitutes an important part of the survey: First, we have to search
close to the detection limit, which makes contamination by noise nonnegilible, and second, even very rare and unlikely contaminants, like
distant supernovae or other transient objects, reach surface densities
comparable to that of the Ly-$\alpha$ candidates, and can be mistaken
as emission line galaxies in our observations of the FPI scans which
are spread over years in some fields.

The first goal of the  spectroscopic follow-up of likely \lya\,
galaxies with 8-m class telescopes is therefore to verify the emission line.
Furthermore, the line shape and the continuum blue- and red-wards of the emission
line could allow us to decide between \lya\, and \OII:
If the line has been verified, but the resolution is not good enough
to identify the emission line, a second step 
is the clear confirmation of the  line using higher resolution grisms,
 in order to see if the line shows the asymetric profile expected for
 \lya, or the double [O\,II]$\lambda\lambda$3726,3729 line with separation $\Delta
 \lambda \cdot (1+z)$, where $\Delta \lambda = 2.75$\am.
If galaxies with bright \lya\, line at $z>4.7$  are confirmed, we plan 
 a detailed study of the corresponding galaxy as  third step.

%

Thus, confirming \lya\, candidates by spectroscopy 
is a slow process.  
Nevertheless, the verification  or non-verification of a line for
every single
\lya\, candidate is  very important
in order to set more robust upper limits to the number counts of
galaxies at high redshift,
 which can set stringent constraints on theoretical models (see section 4).

Spectroscopic follow-up observations of  \lya\, candidates in the 01h-
 and
23h-fields
were obtained in the summer and autumn of 2001,  using  FORS\,2 at the
 VLT.
 The slitmasks contained 1.0\arcsec\, wide slits of
length between 10\arcsec\, and 20\arcsec. The position angle of each
 mask was chosen such that the number of \lya\, candidates  in the 6.8\arcmin\, $\times$
6.8\arcmin\, FORS field of view is maximized.
Regions of the slitmasks not devoted to primary
targets were used to obtain spectra of low metallicity emission line
 galaxies. The results of the observations of  these low metallicity emission line
 galaxies will be presented and discussed in another paper (Maier et
 al. \cite{maier03}, in preparation).
 
Depending on the distribution of candidates, we used two different grisms,
the
lower resolution
300\,I grism, which gives a spectral resolution of about  1.2\,nm,
at 800\,nm, for slitlets 1\arcsec\, wide, and
the 600\,RI grism, which gives a higher spectral resolution of about  0.8\,nm, at 800\,nm,
 for slitlets 1\arcsec\, wide.
%
Three $z \approx 5.7$ \lya\, candidates, 01h-3238, 01h-28090, and
01h-27927, were observed for 215\,min using the 300\,I grism with
FORS2, and one  $z \approx 4.8$ \lya\, candidate, 23h-50707, was
observed for a total of 150\,min   using the
600\,RI grism with FORS2. 
 Fluxes were calibrated using multiple observations of
the spectrophotometric standard stars LTT\,7379, EG\,274 and LTT\,7987
(Hamuy et al. \cite{hamuy92},\cite{hamuy94}).
Seeing varied between 0\arcsec.5 and 1\arcsec.2.

\begin{figure}[!h]
\centerline{\psfig{figure=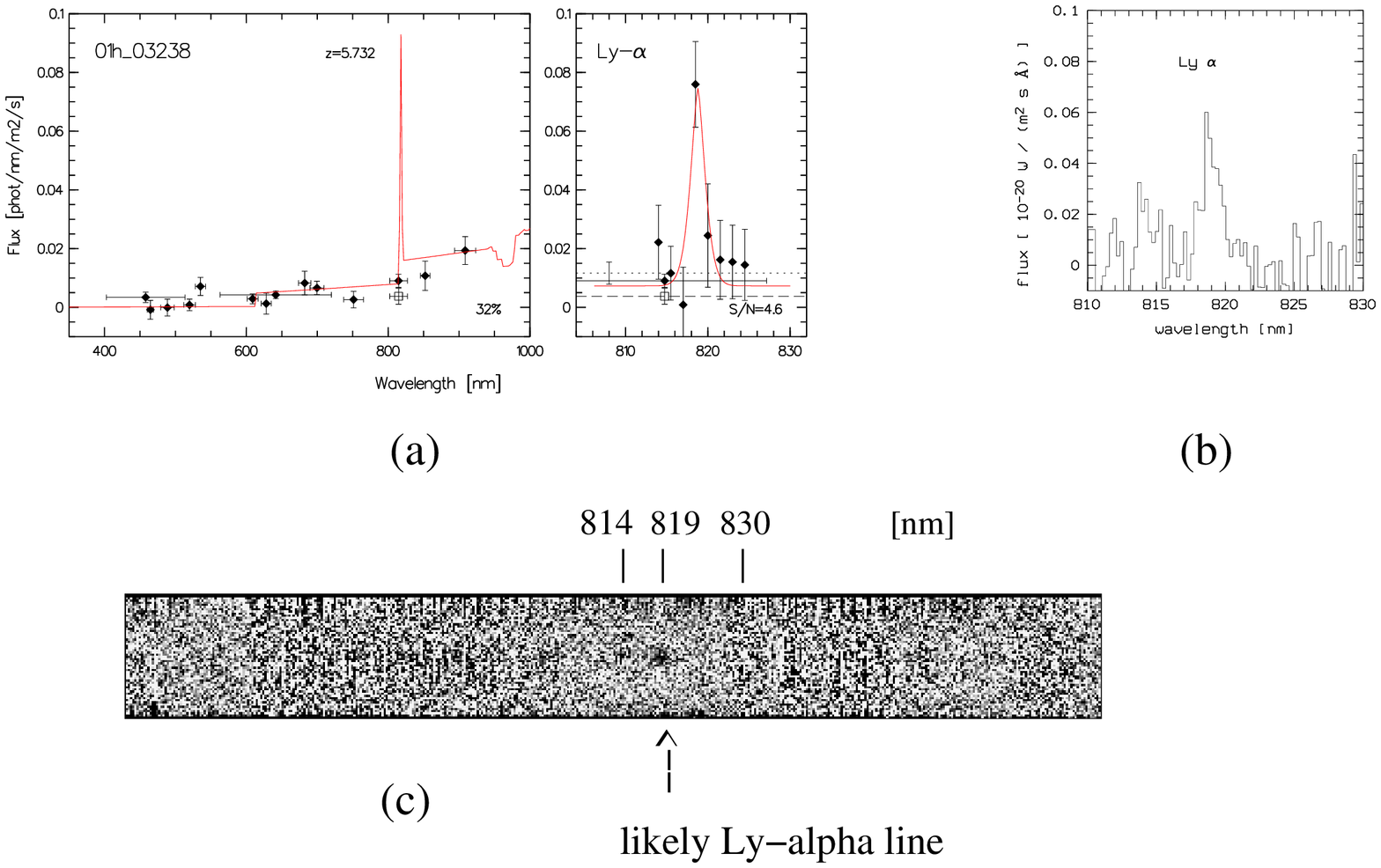,width=16cm,angle=270}}
\caption[01h-3238, a
probable \lya \,galaxy  at $z=5.732$]
{\label{obj3238} \footnotesize  
01h-3238, a
likely \lya \,galaxy  at $z=5.732$.

(a) Photometry
in all 14 optical
CADIS filters fitted by a
  continuum-model  (left panel), and the Fabry-Perot
  measurements with an emission line fit to the observed flux data
  (right panel).

(b) The single emission line, verified with FORS2 at the VLT, using
  the 300I grism.

(c) The two dimensional VLT spectrum.

}
\end{figure}

\begin{figure}[!h]
\centerline{\psfig{figure=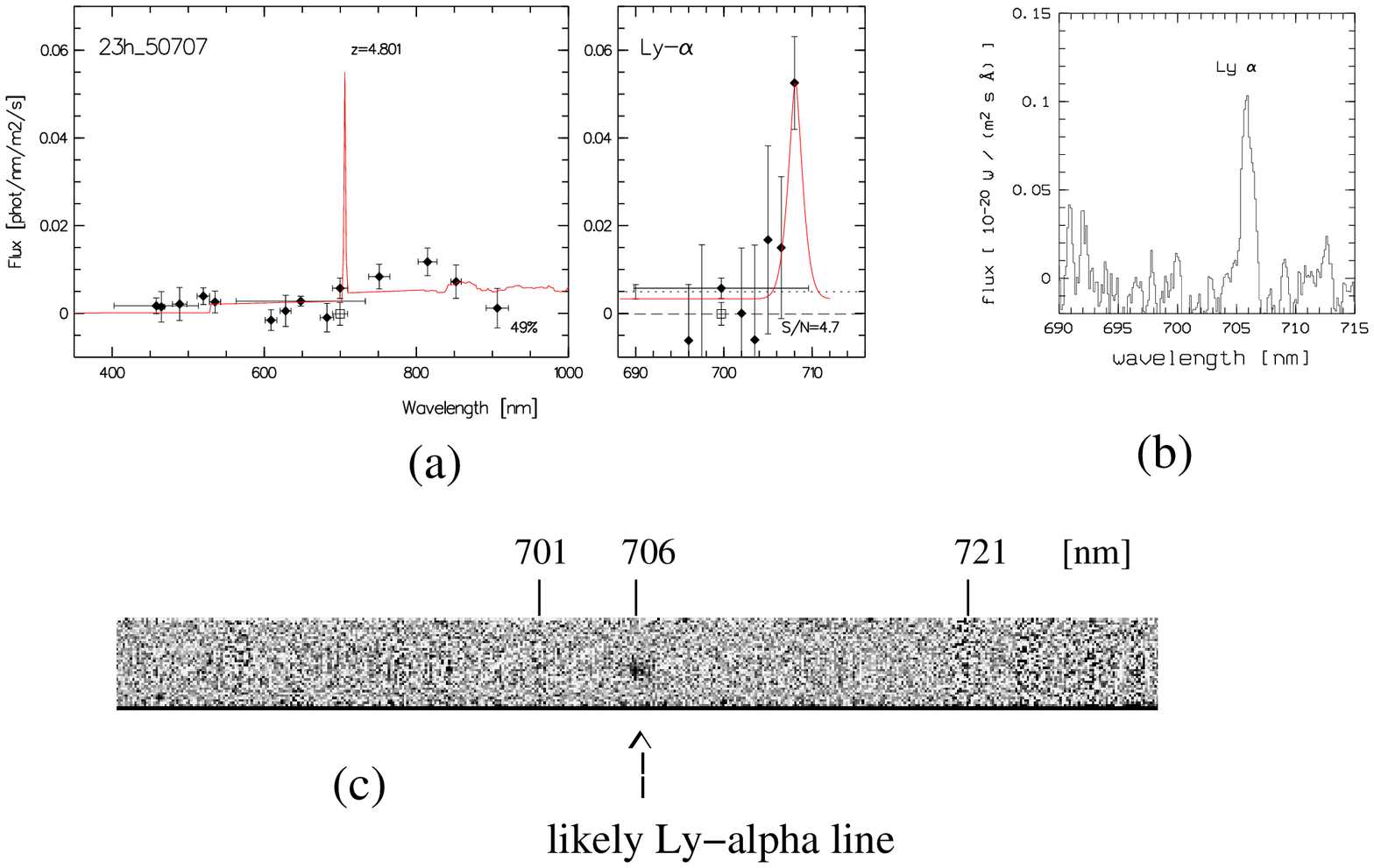,width=16cm,angle=270}}
\caption
[23h-50707,
a probable \lya\, galaxy at $z=4.801$]
{\label{obj50707} \footnotesize  
23h-50707,
a probable \lya\, galaxy at $z=4.801$.

(a) Photometry
in all 14 optical
CADIS filters fitted by a
  continuum-model  (left panel), and the Fabry-Perot
  measurements with an emission line fit to the observed flux data
  (right panel).

(b) The single emission line, verified with FORS2 at the VLT, using
  the 600RI grism.

(c) The two dimensional VLT spectrum.
}
\end{figure}

For two of the four observed Lyman-$\alpha$ candidates we  verified the emission line,
for 01h-3238 at $819.0 \pm 0.3$\,nm, and   for 23h-50707
at $705.7 \pm 0.3$\,nm (see
Figure \ref{obj3238} and Figure \ref{obj50707}).
No continuum and no additional emission lines are seen on the VLT spectra,
and, according to the CADIS measurements, these two objects satisfy the
criteria (1)-(4) for  \lya\, candidates.
Therefore, we conclude that the two emission lines are very likely 
\lya\, lines  of  high redshift galaxies, 01h-3238 at  $z = 5.735 \pm 0.003$ and
23h-50707 at $z = 4.803 \pm 0.003$, respectively.

The line of the other two observed \lya\, candidates, 01h-28090 and
01h-27927, is not seen on the VLT spectra.
These are the \lya\, candidates with the faintest fluxes of our \lya\, candidates list.
One explanation for the non-detection could be that the slit possibly
missed (part of)
the line emitting region of
these two galaxies.
This is indicated by  some emission line objects on the same masks as the
\lya\, candidates which 
show a  spectroscopic determined flux smaller than the CADIS flux.
Thus, the slit width of 1\arcsec.0 used for the VLT observations was possibly too
narrow, and we may have thus measured only  a fraction of the emission
line flux of the \lya\, candidates.
The reason for this is that our astrometry delivers rather accurate
coordinates (better than 0\arcsec.2) for
objects with multiple detections (FP and continuum), but can be off by
more than 0\arcsec.5  for objects detected only in the
Fabry-Perot (\lya\, candidates).
On the other hand, the two \lya\, candidates could be
indeed spurious, since we expect a residual contamination of about
50\% in our list of \lya\, candidates.
The reason for the non-verification of the lines has to be established
by extending the statistics of  spectroscopic follow-up observations.

It should be  noted that we could  verify an emission line (at  $820.9
\pm 0.3$\,nm) also for the
galaxy 01h-4616. 
This object,
which shows no flux in the B filter and in the veto filters, passed
the criteria (1)-(3), but was classified as a probable \OII\, galaxy
at $z \approx 1.2$ according to criterion (4).
 No other
lines are seen on the VLT spectrum, and  a
continuum on both sides of the line is detected.
Therefore,   the emission line is indeed likely to be
\OII\, at $z = 1.202 \pm 0.001$.

We  calculate the absolute \lya\, luminosity of 01h-3238 and
23h-50707 from the observed fluxes measured in the FP scan (since the
spectroscopic observations are subject to slit losses, see above). 
The
Ly$\alpha$ flux is $f$(Ly$\alpha$) = $(4.1 \pm 0.8)
\times 10^{-20}$ W m$^{-2}$  for 01h-3238, and $f$(Ly$\alpha$) = $(4.0 \pm 0.8)
\times 10^{-20}$ W m$^{-2}$  for 23h-50707.
Using  a
 $\Omega_{0}=0.3$,
 $\Omega_{\Lambda}=0.7$,  and H$_{0}=70$\,\kmsmpc\, cosmology, we obtain an
absolute Ly$\alpha$ luminosity of  $L$(Ly$\alpha$) $\simeq (14.5 \pm
3.0) \times 10^{35}$ W for 01h-3238, and 
 $L$(Ly$\alpha$) $\simeq (9.5 \pm
1.9) \times 10^{35}$ W for 23h-50707.
Note that these Ly$\alpha$
luminosities are the highest found among  any  $z \ga 5$ galaxies (luminosities are estimated by using the same
cosmology):
e.g., $3.3 \times 10^{35}$ W for the lensed galaxy HCM 6A at $z=6.56$
     (Hu et al. 2002);
$6.1 \times 10^{35}$ W for SSA22-HCM1 at $z=5.74$
     (Hu et al. 1999);
$3.4 \times 10^{35}$ W for HDF 4-473.0 at $z=5.60$
     (Weymann et al. 1998).
Taking the \lya\, to \Ha\, ratio for Case B recombination and no dust
(Brocklehurst \cite{brockl}), together with the Kennicutt (\cite{kenn}) conversion between
\Ha\, luminosity and star formation rate, 
the derived  star formation rates for the two
\lya\, galaxies  are:  $(14.5 \pm 3.0)M_{\odot}\rm{yr}^{-1}$ for
01h-3238, and  $(9.5 \pm
1.9)M_{\odot}\rm{yr}^{-1}$ for 23h-50707.
Kudritzki et al. (\cite{kudr}) found that the derived star formation rates  depend very strongly on the assumption whether we witness a burst or continuous star formation, the SFR being higher in the case of a starburst.
Therefore, the derived SFRs using  Kennicutt (\cite{kenn}) empirical
formula  for the case of continuous star formation  represent only lower limits for the SFRs. The true values of the SFRs in case of a starburst may be a factor of five higher.


\section{Comparison between observed and theoretical abundances of
  \lya\, emitting primeval galaxies}

The CADIS selection suppresses the contamination in the list of \lya\, candidates to
less than 50\% (see section 2).
As a  consequence, the list of CADIS \lya\, candidates allows us to
set \emph{stringent} upper limits for the density of \lya-emitting galaxies at $z>4.7$.
We can therefore 
discuss the evolution of the population of \lya\, emitting
galaxies from $z=3$ to $z=6$ by comparing the maximum abundance derived
from the list of  CADIS Ly-$\alpha$ candidates  with the results of
 other systematic
surveys at $3<z<6$.

\begin{figure}[h]
\centerline{
\psfig{figure=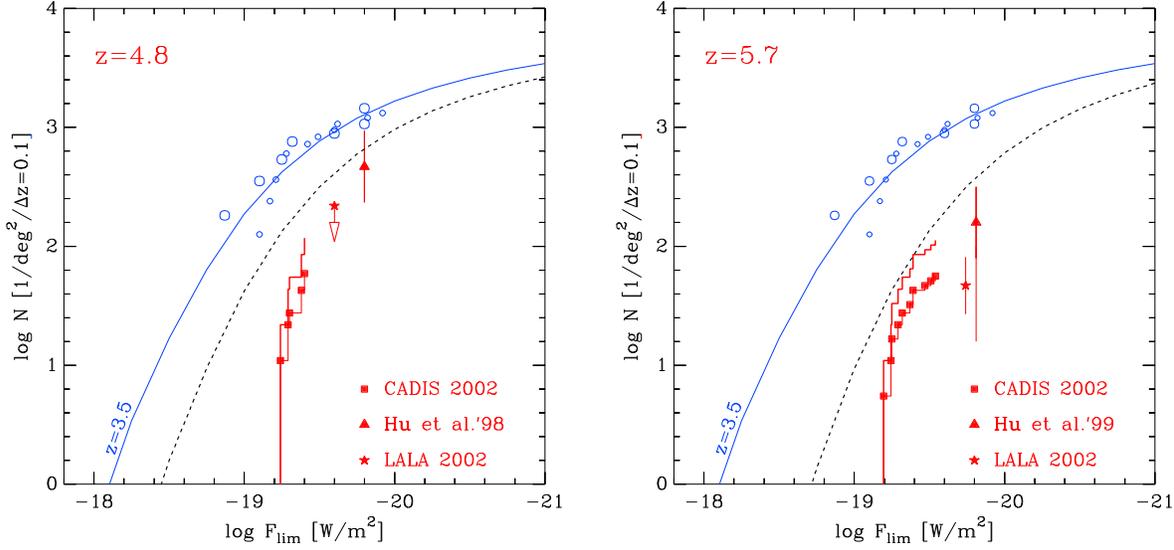,clip=t,width=16.0cm,angle=270}}
\caption[Cumulative number counts of Lyman-$\alpha$-galaxies at
 $z=4.8$  and $z=5.7$]
{
\label{results} \footnotesize  Cumulative number counts of Lyman-$\alpha$-galaxies, $N$ deg$^{-2}$
per $\Delta z = 0.1$, which are
 brighter than a certain observed flux $F_{lim}$,  at
 $z=4.8$ (left panel) and $z=5.7$ (right panel). 
 In the
 cumulative histogram the upper edge 
 represents the number of candidates, while  the  squares take
 into account the likely residual contamination by noise.
 Additional values of the number density  are from  Hu
 et al. (\cite{hu98}) for $z = 4.5$, and from  Hu et al. (\cite{hu99}) at $z=5.7$.
Stars
 show preliminary results from  the Large Area Lyman Alpha Survey (LALA) at $z=4.6$
  (Malhotra \& Rhoads \cite{malh}),
 and at $z=5.7$ (Rhoads et al. \cite{rhoads02}).
 Open circles denote the
 measured abundance of Lyman-$\alpha$-galaxies at $z = 3.5$ through
 emission-line-surveys from Hu et al. (\cite{hu98}, small circles), and
 Kudritzki et al. (\cite{kudr}, large circles).
 A  model luminosity function from Thommes \& Meisenheimer (\cite{thommes03}), fitted to the values   at $z = 3.5$,
 has been converted  to the redshifts searched by CADIS
 assuming \emph{no} evolution between $z = 5.7$ und $z = 3.5$ (dotted
 lines).
%
}
\end{figure}

Figure \ref{results} shows the cumulative number
 counts of  Lyman-$\alpha$-galaxies at $z=4.8$ (left panel) and at
 $z=5.7$ (right panel), presented as the total number
 of galaxies, $N$ per deg$^{2}$ per $\Delta z = 0.1$, which are
 brighter than a certain observed flux, $F_{lim}$.
 In the
 cumulative histogram of CADIS Ly-$\alpha$ galaxy candidates, the upper edge 
 represents the total number of CADIS candidates per deg\,$^{2}$ and per $\Delta z = 0.1$,
 brighter than the respective $F_{lim}$,
 based on the list of \lya\, candidates from Table \ref{listlyman}:
 e.g., at $z=5.7$, in four fields with a total $\Delta \Omega_{total} =
 413$\,arcmin$^{2} \approx 1/9\,\Box^{\circ}$, we find one candidate brighter than $F_{lim}=6.4
 \cdot 10^{-20}\rm{W}\,\rm{m}^{-2}$, two candidates brighter than $F_{lim}=5.7 \cdot
 10^{-20}\rm{W}\,\rm{m}^{-2}$, etc.
 %
 The filled  squares additionally take
 into account the likely residual contamination by noise plus
 unidentified \OII-emitting galaxies, and  therefore represent our
 best-guess upper limit to the abundance of \lya\, galaxies.

 Additional abundance measurements, showed as filled triangles, are    from  Hu
 et al. (\cite{hu98}) at $z = 4.5$, and from  Hu et al. (\cite{hu99}) at $z=5.7$.
 Hu et al.   detected two \lya\, galaxies
 at $z=4.52$ in a 24 arcmin$^{2}$ field, and one \lya\, galaxy at $z=5.74$ using
 narrowband observations in a 30 arcmin$^{2}$  field.
 Stars
 indicate preliminary results from  the Large Area Lyman Alpha Survey
 (LALA)  at $z=4.6$
  (Malhotra \& Rhoads \cite{malh}) and at $z=5.7$ (Rhoads et
 al. \cite{rhoads02}).
Malhotra \& Rhoads (\cite{malh}) found 157 \lya\, \emph{candidates} at
$4.37<z<4.57$  in one field of 0.36\,deg$^{2}$.
The arrow in Fig.\,\ref{results} (left panel) indicates the upper
 limit of the number density of \lya-emitters based on the list of these candidates.
 Rhoads et al. (\cite{rhoads02}) reported 18 \lya\, \emph{candidates} at $z\approx
5.7$ in one field of $\sim 0.2$\,deg$^{2}$, of which three (out of four
 observed) candidates   
 have been verified spectroscopically. Rhoads et al. infer
 a number density    $39<N<54$ of \lya-emitters per  deg$^{2}$
 per $\Delta z =0.1$ down to their detection limit (see
 Fig.\,\ref{results}, right panel).
 For comparison, we show in Fig.\,\ref{results} the
 measured abundance of Lyman-$\alpha$-emitting galaxies at $z = 3.5$ through
 emission-line-surveys from Hu et al. (\cite{hu98}, small circles), and
 Kudritzki et al. (\cite{kudr}, large circles; converted from $z=3.1$).
 
 In order to parametrize the observed number counts at $z=3.5$, we use
 the model by Thommes \& Meisenheimer (\cite{thommes03}): It basically
 assumes (i) that \lya-galaxies mark the onset of star-formation in
 present-day spheroids, (ii) a rather short-lived \lya-bright phase
 due to rapid dust formation ($\sim 10^{8}$\,yrs), and (iii) a mass
 dependent formation history as given by the peak formalism.  
 In order to fit the abundance at $z=3.5$ we used model parameters
 which set the peak of \lya-emission for the bulge of the Milky Way
 around $z\simeq 6$. The normalization at $z=3.5$ is set by the
 abundance of present day spheroids together with a free parameter
 $-$ the duration of the \lya-bright phase. For the present work, we
 decided not to fine tune the model in a way that a consistent
 description of the abundances at all redshifts $3.5 \leq z \leq 5.7$
 is obtained. This will be presented in a forthcoming paper. Here we
 rather would  like to demonstrate a robust qualitative statement
 $-$ namely that \lya-galaxies are less common at $z=5.7$ than at
 $z=3.5$. To this end we simply \emph{shift} the model function for
 $z=3.5$ to higher redshift by taking into account both the higher
 luminosity distance (shift to the right), and the smaller comoving
 volume $\Delta V/ \Delta z$ (shift to bottom). This ``no-evolution
 model''    is shown by the dotted line in Fig.\,\ref{results}.

It is obvious from Fig.\,4 that the CADIS upper limits both at $z=4.8$
and $z=5.7$ fall short of such a non-evolution model, while at the faint
end,  F$_{\rm{line}} \leq   10^{-20}\,\rm{Wm}^{-2}$, the present
results could still be compatible with no evolution.

Since our statistics at  $z\simeq 4.8$ is extremely limited, we will
focus any further discussion on redshifts around $z=5.7$.

While the number counts of Hu et al. 
at  $z=5.7$ (a single
galaxy  found by Hu et al. \cite{hu99}) still seem compatible with the
no-evolution model, the abundance of \lya-galaxies from LALA and the
limits given by the CADIS candidates fall significantly short of the
no-evolution prediction.
The under-abundance of \lya\,
 emitting galaxies at $z=5.7$ compared to $z=3.5$ becomes even more obvious if one considers that the
noise distribution of our survey predicts  residual contamination
of about 50\%.
Nevertheless, before drawing definite conclusions, one should consider
whether selection effects could lead to an under-estimation of the  number of
\lya-emitting galaxies.
%
%
Potential influences on the upper limits
 include: objects classified
 as probable \OII\, galaxies for which the emission line is actually 
 \lya, the fainter surface brightnesses of high redshift objects,
 and the large-scale structure. The weakly confined bright end of the
 luminosity function at $z=3.5$
 may limit the comparison at the bright end. Therefore, we have to
 discuss how robust the CADIS upper limits are.

The distinction between a  galaxy with  \OII\,
emission, but  no flux in the
CADIS B filter, and a Lyman-$\alpha$-galaxy is
ambiguous (see step (4) in Section 2).
Therefore, some of the probable \OII\, objects may turn out to be
 \lya\, galaxies, and some presumed \lya\, galaxies may turn out to be
 \OII-emitting galaxies.
However,
any asymmetry in this mutual contamination has already been accounted
for by assuming a conservative estimate of 50\% contamination of the
\lya\, candidates sample.


Surface brightness depends on the
apparent size of an object at redshift $z$, which 
does not change much between $z=3.5$ and $z=5.7$, and on the luminosity distance. 
Since the model   luminosity  function is converted to $z=5.7$
taking the luminosity distance into consideration, most of the effect
of the $(1+z)^{-4}$ dimming towards 
higher redshift has  already
been accounted for.

Large-scale structure can influence galaxy counts even at high
redshifts (Steidel et al. \cite{steidel98}).
Since we derive 
the number counts 
of \lya\, galaxies at $z\approx 5.7$ by combining
four CADIS fields (with a total area  about 14 times the size of the field searched by Hu et al. \cite{hu98}), large scale structure should average out. 
Nevertheless, it should be kept in mind that results from small fields, like those searched by Hu et
al. (\cite{hu98}, \cite{hu99}),   could be affected by
large scale structure.

Two points should be noted  about the bright end of the luminosity function at $z
\approx 3.5$.
First, the \emph{observed}  \lya\, line fluxes for  the most luminous galaxies at $z \approx 3.5$ found
by  Hu et al. (1998) and Kudritzki et al. (2000) are on the order of
$10^{-19}\rm{W}\,\rm{m}^{-2}$.
Taking into account  the higher luminosity
 distance at   $z=5.7$ (compared to $z=3.5$) these galaxies
 would have apparent line fluxes of 
 $\sim 3
 \times 10^{-20}\rm{W}\,\rm{m}^{-2}$ at  $z=5.7$ (using  a
 $\Omega_{0}=0.3$,
 $\Omega_{\Lambda}=0.7$  and H$_{0}=70\,$\kmsmpc\, cosmology);
i.e., galaxies with intrinsically  bright \lya\, emission,
corresponding to  observed line
fluxes  $> 3 \times 10^{-20}\rm{W}\rm{m}^{-2}$ at $z \approx 5.7$
$-$ as searched by
 CADIS $-$  have not yet been found  in the small fields searched at $z \approx 3.5$.
 Thus, 
the shape of the bright end  of the luminosity function relies mainly
on our model.
Second, the number counts at the bright end of the luminosity
function at
$z=3.5$, resulting from the   brightest galaxy detected  by Kudritzki et
al. (2000, large circles),  are higher than the results of Hu  et
al. (1998, small circles).
Since large-scale structure can influence galaxy counts even at high
redshifts (Steidel et al. 1998), the  Kudritzki et al. number counts
could be biased, because Kudritzki et
al.  searched  only one field,
whereas Hu  et al. survey goes over two fields.
Disregarding the  brightest galaxy from Kudritzki et al.
would result in a
steeper luminosity function at the bright end at $z=3.5$ and thus
lessen the discrepancy at $z=5.7$.


In summary, we conclude that the luminosity function of \lya-bright
galaxies declines between $z=3.5$ and $z=5.7$. This implies that the
peak of the \lya-bright phase $-$ i.e., the first formation of massive
stars $-$ is reached after $z=6$. Although, in principle, both
luminosity evolution or density evolution could account for this
finding, we only can establish the decrease in density around $3 \times
10^{-20}\rm{W}\rm{m}^{-2}$, as the bright end of the \lya\, luminosity
function is still not  determined well enough
at $z=3.5$.

The onset of massive star formation should contribute a substantial
fraction of the UV photons which reionize the universe. Moreover,
both bright starbursts (SFR $\geq 10\,
\rm{M}_{\odot}$/yr) and the formation of supermassive black holes
indicate a strong concentration of baryons within the centers of the
most prominent density peaks. Therefore, our result seems to confirm
additionaly the finding from the analysis of the HI absorption in the
spectra of SDSS quasars at $z>6$ (Fan et al. \cite{fan02}), that the reionization
of the universe did not happen long before $z=6$.


%

\clearpage
\newpage


\newpage

\end{document}